# Evry Léon Schatzman

Jean-Pierre Luminet
Laboratoire Univers et Théories, CNRS-UMR 8102,
Observatoire de Paris, F-92195 Meudon cedex, France
E-mail: jean-pierre.luminet@obspm.fr

This article describes the life and work of French astrophysicist Evry Schatzman (1920-2010). He was a pioneer in the study of white dwarfs during the 1940s and was one of the proponents of the wave heating theory of the solar corona. He made important contributions to the fields of internal stellar structure, novae, mechanisms of acceleration of cosmic rays, the role of turbulent diffusion in stellar evolution and its consequences for the lithium abundance, and the rate of solar neutrinos. Schatzman is mostly recognized as the creator of the French school of theoretical astrophysics. Although he was not the first theoretician of astrophysics in his country, he was the first to have felt the need for a rapid development of this subject in France, and the first to teach it and to guide the path of many young researchers. Many of them became involved, and some leaders, in space science.

Born : Neuilly-sur-Seine, France, 16 September 1920 Died : Paris, 25 April 2010

There is hardly a field of theoretical astrophysics that was not addressed by Evry Schatzman. His father, Benjamin, was a dentist born in Tulcea, Romania, who emigrated at a young age with his family in Palestine during the First Aliyah (1882-1903). As an adult he settled in France. Evry's mother, Cecile Khan, was the daughter of a secretary of the Israelite Consistory in Paris.

Evry began to study fundamental physics at the Ecole Normale Supérieure of Paris in November 1939. After the German invasion of France, Schatzman fled occupied France because of the Vichy government's anti-Jewish laws, and arrived in Lyon in January 1942 to continue his studies. He worked there for a year and met his wife, Ruth Fisher. The newlyweds were able to take refuge from the turbulent events of that period at the Observatoire de Haute-Provence (OHP), thanks to the help of its courageous director, Jean Dufay (1896-1967), and the astronomer Charles Fehrenbach (1914-2008), who took in charge several people of the Jewish faith during the German occupation. However, Evry Schatzman was never to see his father again, who fell victim of a roundup in December 1941 and died at the Auschwitz concentration camp.

Schatzman remained at the OHP until the end of the hostilities, under the false identity of Antoine Emile Louis Sellier. At the library of the OHP, he read the proceedings of a symposium organized by the German-born American astronomer Walter Baade (1893-1960) on white dwarfs, the compact remnants of moderately massive stars. Excited by this book, Schatzman developed a description of the structure of these stars, which would be the subject of his thesis.

After finally returning to the Ecole Normale Supérieure, rue d'Ulm in Paris at the end of 1944, he got the Agrégation of Physics (the highest teaching diploma in France) and began working at the Centre national de la recherche scientifique (CNRS) in the fall of 1945. He defended his doctorate thesis in March 1946 in front of a jury chaired by the Nobel prize winner in Physics, Louis de Broglie.

From that moment on, Schatzman's career path was straightforward. He worked as a researcher and supervisor first at the Institut d'Astrophysique de Paris (IAP), where he published influential papers about supernovae and the internal structure of white dwarfs. In 1947 he worked at Copenhagen University with Bengt Strömgren (1908-1987), who oriented him towards the study of white-dwarfs atmospheres. Schatzman realized that the atmospheres of white dwarfs should be gravitationally stratified, with hydrogen on top and heavier elements below, and modeled pressure ionization in their atmospheres, which help to explain their spectrum.

Schatzman was then invited for a year at Princeton University, where the astrophysicist Lyman Spitzer (1914-1997) and the astronomer Martin Schwarzschild (1912-1997) were his mentors. There he studied stellar atmospheres and proposed the mechanism of heating of the solar corona and chromosphere by shock waves.

Schatzman returned to France in 1949 as a researcher at IAP and also gave astrophysics courses at the Sorbonne, along with André Danjon, who occupied the chair of astronomy (the only one at that time in France).

In 1954 Schatzman himself became the first chair holder for Astrophysics created in France at the faculty of sciences of Paris. From 1949 to 1967 he also taught regularly at the Free University of Brussels (ULB). In 1959 he began the teaching of theoretical astrophysics. He also created and chaired two DEA (Diplôme d'Etudes Approfondies) devoted, respectively, to astrophysics and the physics of ionized plasmas. From now on Schatzman directed the majority of astrophysics doctoral theses defended in France - and often overseas - until his own students took over. Many of his students were women, who became a significant contribution to the large fraction of female astronomers in France from the 1960s onward.

In 1969 Schatzman moved to the Paris-Meudon Observatory, where he initiated the creation of the Laboratoire d'Astrophysique de Meudon (LAM), which he directed for several years and where many innovative space research ideas found their roots, such as for example, the detection of extra solar planets from space, being now performed by the CoRoT and Kepler missions.

In 1976 Schatzman moved to Nice Observatory, where he became a full-time researcher nearly until the end of his official career. He was also a researcher at the University of California at Berkeley between 1984 and 1988. As he retired in the fall of 1988, he finally returned to Meudon Observatory to have the leisure to develop his ideas, and he stayed there as an active emeritus director of research, continuing his works on the physics of astrophysical environments.

Schatzman's career was studded with numerous accolades and awards. He received the Prix Paul et Marie Stroobant of the Royal Academy of Sciences of Belgium and the Prix Robin of the French Physical Society in 1971, the Prix Jules Janssen of the French Astronomical Society in 1973, the Holweck Award in 1985, and the Gold Medal of the CNRS in 1983 – the highest scientific honor awarded in France and the distinction of which he was proudest. In 1985 he was elected at the French Academy of Sciences.

Evry Schatzman's research continues to provide important arguments and driving ideas to theoretical astrophysics. This is true of his concepts on the diffusion of elements in astrophysical environments, the thermonuclear origin of stellar energy, the propagation of seismic pressure and gravity waves through the layers of the Sun, or the evolution of stars. He had rare creative intuition and an acute sense of what phenomena are important or significant.

He was also one of the first to propose detecting antimatter from space, a very challenging experiment indeed, which wasbegun 1 year after his death. The Alpha Magnetic Spectrometer (AMS) is a particle physics experiment module designed to search for various types of unusual matter by measuring cosmic rays. The launch of the Space Shuttle Endeavour flight carrying AMS took place on 16 May 2011, and the spectrometer was installed on the International Space Station on 19 May 2011. In July 2012, it was reported that AMS had recorded over 18 billion cosmic ray events since its installation.

Schatzman was not a scientist who got lost in abstractions. Concerned about the changing world, still affected by the indelible memories of the years of war, naturally he turned to political commitments. Although a member of the French Communist Party in 1946, which earned him insults from some of his colleagues and closed the doors of the United States to him for many years, he left the party in 1959, shocked by the excesses of Stalinism. But he retained a strong sense of activism that led him to become a union activist. He was also involved in a philosophical action-oriented school of thought and, from 1970 to 2000, headed the Rationalist Union, an association that had been set up by Henri Roger and physicist Paul Langevin.

At its creation in 1930, the Rationalist Union promized to publicize the great discoveries of contemporary science, the problems posed by these discoveries, and the spirit and methods of scientific work. Schatzman provided impetus to the association given the fact that today, the questions about science are multiplied and with them so are the questions on scientific culture and the teaching of science. Reading some of the editorials he wrote regularly for the *Cahiers rationalistes*, we cannot to be struck by heart and tenacity with which he has led, over the years, this fight for reason and science.

Schatzman often worried the unease between science and the public. He attributed it largely to the form of education that emphasizes the divorce between scientific culture and what is generally called just culture. As he wrote, we must distinguish between teaching *what is* science and teaching science. Rationalist formation is a formation of the judgment, an education of the mind. It is the opposite of dogmatic education. The meeting of reason and democracy is the way of the expansion of democracy and the development of the reason.

Evry Schatzman published several books in French and in English. With his colleague Jean-Claude Pecker, he wrote the astrophysics textbook *Astrophysique Générale*, which contributed greatly to the formation of French students in astrophysics. In his popular science books such as *Sciences et société* (1971) and especially *La science menacée* (1989), he denounced, on the one hand, the external threats from those who consider the scientific realism as "destroyer of the world of sensitivity and emotion...", but also of those who confuse the discoveries and their applications. Among external threats, on the other hand, he condemned scientism, according to which scientific progress would result *ipso facto* in social progress, as well as a movement of thought called relativism, which more and more refuses the findings of scientific knowledge.

**Selected References**

Jean-Claude Pecker, *Evry Schatzman, astrophysicien*, Le Monde n°20300, 30 april 2010, page 23.
Jean-Claude Pecker, *Evry Schatzman Obituary*, http://iau-c35.stsci.edu/News/index.html

## Selected works

*Origine et évolution des mondes*, Paris: A. Michel, 1957; translated into English by Bernard and Annabel Pagel as *The origin and evolution of the universe*, New York: Basic Books, 1965.

*White Dwarfs*, Amsterdam: North-Holland, 1958.

(with Jean Claude Pecker) *Astrophysique Générale*, Paris: Masson, 1959.

*Science et Société*, Paris : Robert Laffont, 1971.

*Les Enfants d'Uranie : à la recherche des civilisations extraterrestres*, Paris : Le Seuil, 1986.

*Le Message du photon voyageur*, Paris : Belfond, 1987.

*La Science menacée*, Paris : Odile Jacob, 1989.

*Our Expanding Universe*, New York: McGraw–Hill, 1992.

(with Françoise Praderie) *Les Etoiles*, Paris: Paris Interéditions et ed. du CNRS, 1990. Translated into English by A. R. King as *The Stars*, Berlin: Springer, 1993.

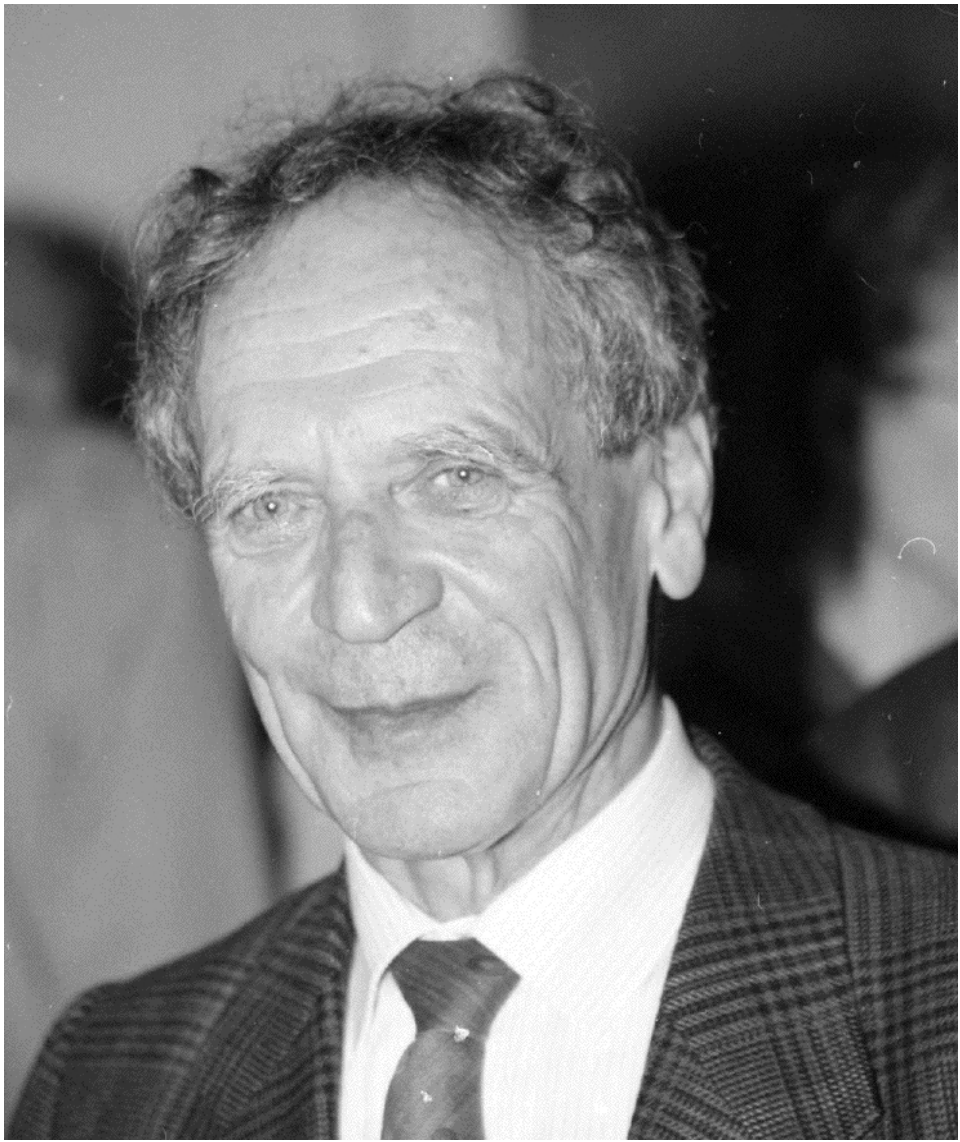

A portrait of Evry Schatzman